\documentclass[apj]{emulateapj}

\usepackage{graphicx}

\shorttitle{A method to search for correlations of UHECR masses with LSS}
\shortauthors{A.A. Ivanov}

\begin{document}

\title{A method to search for correlations of ultra-high energy cosmic ray masses\\
with the large scale structures in the local galaxy density field}

\author{A.A. Ivanov}
\affil{Shafer Institute for Cosmophysical Research and Aeronomy,
31 Lenin Avenue, Yakutsk 677980, Russia}
\email{ivanov@ikfia.ysn.ru}

\begin{abstract}
One of the main goals of investigations using present and future giant extensive air shower (EAS) arrays is the mass composition of ultra-high energy cosmic rays (UHECRs). A new approach to the problem is presented, combining analysis of arrival directions with the statistical test of the paired EAS samples. An idea of the method is to search for possible correlations of UHECR masses with their separate sources, for instance, if there are two sources in different areas of the celestial sphere injecting different nuclei, but fluxes are comparable so that arrival directions are isotropic, the aim is to reveal a difference in the mass composition of CR fluxes.
The method is based on a non-parametric statistical test -- the Wilcoxon signed-rank routine -- which does not depend on the populations fitting any parameterized distributions. Two particular algorithms are proposed: first, using measurements of the depth of EAS maximum position in the atmosphere; and second, relying on the age variance of air showers initiated by different primary particles.
The formulated method is applied to the Yakutsk array data, in order to demonstrate the possibility of searching for a difference in average mass composition of the two UHECR sets, arriving particularly from the supergalactic plane and a complementary region.
\end{abstract}



\section{Introduction}
The origin of ultra-high energy cosmic rays (UHECRs) is a long-standing challenge for astrophysics. The energy spectrum of particles constituting cosmic rays (CRs) is measured up to and slightly above $10^{20}$ eV (= 100 EeV) \citep{HiRes08,PAO,CRIS}, but no evidence has been revealed till now, either of the sources or the sort(s) of highest energy particles, owing mainly to the arrival direction distribution, which is nearly isotropic~\citep{Cronin, Grieder}. However, some hints have been found recently of the possible correlation of UHECR arrival directions with nearby active galactic nuclei (AGN) at energies $E>56$ EeV~\citep{AGN}.

Disputable estimates of the mass composition of the highest energy CRs were given, based on the average depth of EAS maximum, $x_m$, and variance, $\sigma(x_m)$, measured by the Pierre Auger Observatory (PAO), the High Resolution Fly's Eye (HiRes), the Telescope Array (TA), and the Yakutsk array \citep{AWG}.
One possible explanation of the diverging results may be the different average masses of UHECRs, observed in different fields of view of the instruments.

In this paper, another approach is used to search for the possible correlation of UHECR masses with their sources, based on extensive air shower (EAS) observables, namely $x_m$ and the shower age, $\tau$, varying with the mass of primary particles. Here, `age' means the stage of the cascade development at the detector level, $x_0$: $\tau=x_0\sec\theta/x_m$, where $\theta$ is the inclination angle of the shower axis \citep{IJMP}. This method is convenient for revealing different primary particles initiating EAS, rather than to search for excess flux from candidate sources of UHECRs.

The paper is structured as follows: the new method is described in Section 2, where two algorithms are formulated using the depth of EAS maximum and the shower age measurements, and a non-parametric statistical test is applied to distinguish a pair of samples. In Section 3, the method is tested and applied to the Yakutsk array data observed in the energy range above $1$ EeV. Conclusions are given in the final section.


\section{Searching for a difference in mass composition of UHECRs arriving from complementary celestial regions}
The method is aimed at the possible differences in the mass of UHECRs arriving from different regions in the celestial sphere. For example, in models of CR acceleration by shocks in AGN relativistic jets the mechanisms are proposed where the maximal energy of CRs is proportional to the particle charge, and where protons could be accelerated up to energy $\sim100$ EeV \citep{Berezhko}, and iron nuclei to energy $\sim3000$ EeV (in Cen A, \cite{Honda}). Observable particle energies should be reduced due to fragmentation and energy loss in the intergalactic medium and in the Galactic wind. However, in the Cen A case, the object is nearby ($<5$ Mpc), so that the energy is not crucially degraded.

By selecting EASs with arrival directions in the vicinity of Cen A in contrast with the complementary area where, presumably, protons dominate, one can reveal the fraction of heavy nuclei, if the model of \cite{Honda} is applicable. An obstacle is deflections of particles in magnetic fields -- it was shown that protons of GZK energy ($\sim40$ EeV) are deflected a few degrees coming from any source within 100 Mpc \citep{Sommers}. Deflections of iron nuclei from Cen A would be greater by $\sim30\%$, so that the energy of particles detected should be well above the GZK energy.

In the transition region between galactic and extragalactic components of CRs, the method can be used to verify the difference in mass composition of the two components comparing, for example, equatorial and polar regions in galactic coordinates.

The method can be regarded as an extension of the matter tracer model proposed by \cite{Tinyakov}, for the case where different particles are supposed to be generated in UHECR sources.

Our first task is to test the null hypothesis; that is, that there is no difference in mass composition between two regions. We have to compare two distributions of some EAS observables sensitive to composition of the primaries in the given energy range, in order to decide - whether there is a significant difference or not.

If the difference exceeds experimental errors, then the only cause should be the mass composition in UHECR samples\footnote{We do not consider the case of different CR energy spectra from different celestial regions.}. So, the next task would be to evaluate the most probable value and confidence interval of the mass difference. In this paper we focus on the first task.


\subsection{Non-parametric statistical test of data samples}
The distributions of measured EAS parameters are not described usually by the normal distribution, and moreover, have a specific form in each particular case, so that the general approach to the statistical test of the data samples is preferably non-parametric. The meaning of the term refers here to distribution-free methods, which do not rely on assumptions that the data are drawn from a known probability distribution. Non-parametric methods can be used, for instance, in studying samples of EAS data where certain assumptions cannot be made about the original population. An additional advantage of the approach is that, even in the case when the use of parametric methods is justified, non-parametric methods are easier to use, and  leave less room for improper use and misunderstanding.

In this paper, a Wilcoxon test for a pair of samples is used in data analysis. This test is one of the widely known, non-parametric significance tests (open-access description is given by \cite{Wilc}). It is useful for deciding whether the two samples of observations belong to the same original distribution: the null hypothesis, $H_0$, is that the two samples are drawn from a single population.

Hereafter, we are going to consider pairs of matched samples, $S$ and $T$, containing an equal number of measurements, $N$. In this case, the test is called the `Wilcoxon signed-rank test' (WSRT), in contrast to the case with independent samples, named the `Wilcoxon rank-sum test', or Mann-Whitney U test. As an example of paired samples $S$ and $T$ in the case of cosmic rays, we can consider two samples of $x_m$ measurements in EAS detected in Summer and in Winter  with the same array, with equal energies in pairs of events. Independent samples are those measured in the same energy interval, but without equality in pairs.

The test procedure is as follows. Excluding all pairs in samples where observed values are equal, $S_i=T_i$, we reduce $N$ to the number of pairs with unequal measurements. Rank the differences $|S_i-T_i|$ in ascending series, where rank is assigned as an item number in a series. The WSRT statistic, $W$, is then a sum of signed ranks\footnote{Sign is +1, if $S_i>T_i$ and is -1, if $S_i<T_i$.}; $|W|\leq0.5N(N+1)$. Under $H_0$ the distribution of $W_0$ is symmetric with $\overline{W}=0$ and $\sigma_W=\sqrt{N(N+1)(2N+1)/6}$.
There are tabled values of the ratio $z_0=(W_0-0.5)/\sigma_W$ with which to compare the resultant statistic.

The probability $P_{crit}(z>z_0)$ is a measure of the significance of deviation from the expected value under $H_0$. In the following, $P_{crit}=0.01$ is assumed as the critical value for rejecting the null hypothesis.

\begin{figure}[t]
\includegraphics[width=\columnwidth]{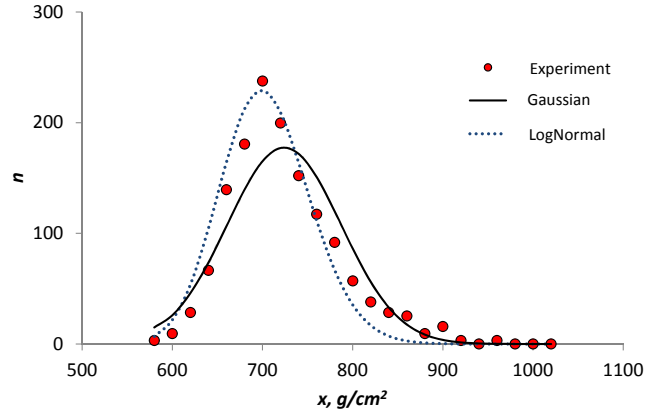}
\caption{Distribution of $x_m$. Experimental data are from Figure 3 in \citep{PAOxm}: 1407 events in $1<E<1.26$ EeV, $\overline{x_m}=713$ g/cm$^2$, $\sigma(x_m)=56$ g/cm$^2$.}
\label{fig:Xm}
\end{figure}


\subsection{Statistical power of the Wilcoxon test}
In this section, one of the composition-sensitive EAS observables, i.e. the depth of shower maximum, will be used to find out the efficiency of the WSRT in the case of EAS samples. Different primary nuclei initiating EAS result in different $x_m$ of the shower (see, for example, recent review by \cite{Kampert}). The question is: What is the minimal number of measurements needed to reject the null hypothesis at the significance level $P_{crit}$, having given difference, $\Delta x_m$? In other words, if an alternative hypothesis, $H_1$, is true, that the two samples are drawn from different distributions with a given difference in average values, $\Delta x_m$, what is the probability to rule out the null hypothesis having samples of size $N$? Inverting the problem, we have to find the sample size needed to reject $H_0$ at a confidence level 99\%.

It is known that, in general, the ratio of Wilcoxon test's efficiency to Student's $t$-test is $3/\pi$, but if the distributions are far from the Gaussian and for large sample sizes, the Wilcoxon test can be considerably more efficient than $t$-test \citep{Waerden}.

In Figure \ref{fig:Xm} the distribution of $x_m$, measured by the PAO Collaboration, is given \citep{PAOxm}. It is undoubtedly non-Gaussian\footnote{The $\chi^2$-deviation from the observed numbers is 168.6 for the normal approximation in 17 intervals, and is 1426 for the lognormal approximation, while less than 32 expected at the significance level $P_{crit}$ for good fit.}. So, we will use another variable, $x_G$, with normal distribution, which has the same average value and RMS deviation in the same energy interval $1<E<1.26$ EeV, in order not only to determine the efficiency of the Wilcoxon test, but to compare it with a Student's $t$-test, known to provide an exact test for the equality of the means of two normal populations with equal variances.

Let the samples $S$ and $T$ of the size $N$ be from normal population, $\overline{x_S}=713$ g/cm$^2$, $\overline{x_T}=\overline{x_S}+\Delta x_m$, $\sigma(x_S)=\sigma(x_T)=56$ g/cm$^2$. Samples are paired, so that $S_i$ and $T_i$ are for showers of the fixed difference in energy of the primaries. Varying the corresponding difference of the mean depths, $\Delta x_m$, one can calculate the minimal sample size, $N_{min}$, needed to distinguish samples $S$ and $T$ by the two tests.

\begin{figure}[t]
\includegraphics[width=0.9\columnwidth]{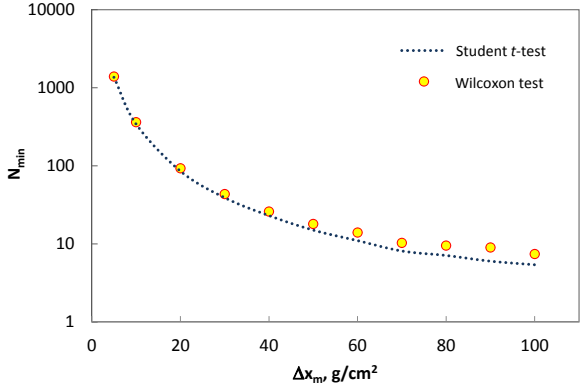}
\caption{Comparison of the efficiency of Wilcoxon and Student tests. $N_{min}$ is the minimal sample size required to reject the null hypothesis, when an alternative hypothesis is true with a given difference, $\Delta x_m$, in paired EAS samples.}
\label{fig:Comparison}
\end{figure}

Results of the Monte Carlo simulations are shown in Figure \ref{fig:Comparison}. The conclusion is that the ratio $N_{min}^{Student}/N_{min}^{Wilcoxon}$ is approximately 0.7 at large differences $\Delta x_m$, and approaching 1 at minimal $\Delta x_m$. It means that the $t$-test is more efficient at small sample sizes, i.e. $N_{min}^{Student}$ can be reduced to $\sim70\%$ of Wilcoxon test's number with equal statistical power. However, in our case, asymptotic efficiencies of the tests, defined as the limit of the efficiency as the sample size grows, are equal. The ratio is valid in the case of normal distributions, where the $t$-test is applicable.

\subsection{A method based on the measurement of the depth of EAS maximum}
In the UHECR domain, the shower maximum is directly observable with air fluorescence detectors (FDs). Collaborations working at HiRes \citep{HiResXm}, PAO \citep{PAOxm} and TA \citep{TAxm} have measured $\overline{x_m}$ as a function of energy, to estimate the mass composition of CRs. In future detectors such as JEM-EUSO, Auger Next, etc, the fluorescence technique is planned to be used, including $x_m$ measurement, along with other developments. In all these experiments, the proposed method can be applied to search for correlations of UHECR masses with their sources.

\begin{figure}[t]
\includegraphics[width=\columnwidth]{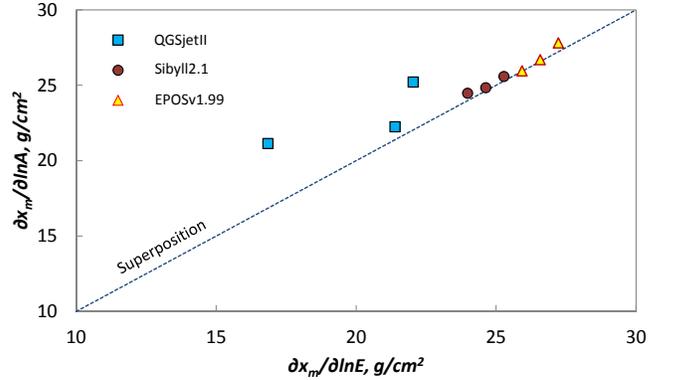}
\caption{Relation of elongation rates in EAS initiated by nuclei. CORSIKA simulation results (taken from the review of \cite{Kampert}) for models are shown by three points calculated at $E=0.1/1/10$ EeV. Superposition approximation results in $|\partial x_m/\partial \ln E|=|\partial x_m/\partial \ln A|$.}
\label{fig:ERdLnA}
\end{figure}

A relation between $x_m$ and CR mass, $A$, is obvious in a superposition approximation, where an air shower, initiated by the nucleus of mass $A$, is treated as a superposition of $A$ nucleon initiated showers of energy $E/A$. So, the depth of a shower maximum is \citep{Linsley}: $x_m=x_{18}+D_{ER}\ln(E/A)$, where $D_{ER}$ is elongation rate, $E$ is in EeV. Averaging it over the distribution of CR energy and mass, $J(E,A)$, we have
\begin{equation}
\overline{x_m}=x_{18}+D_{ER}\overline{lnE}-D_{ER}\overline{ln A}.
\label{Eq:avER}
\end{equation}
Assuming $A$ as the slowly changing function of energy, we can use the relation
\begin{equation}
\Delta\overline{x_m}=D_{ER}\Delta\overline{\ln A}
\label{Eq:Dxm}
\end{equation}
in a narrow energy interval.

The accuracy of the estimations (\ref{Eq:avER}) and (\ref{Eq:Dxm}) can be evaluated in comparison with the model simulation results. Monte Carlo codes such as CORSIKA \citep{CORSIKA}, with implemented hadronic interaction models, give a more realistic description of the cascade than the superposition approximation. In Figure \ref{fig:ERdLnA} a comparison is presented of the results of CORSIKA simulations (QGSjetII, Sibyll2.1 and EPOSv1.99 are implemented \citep{Kampert}) with equation (\ref{Eq:avER}). The difference of QGSjet results from that of equation (\ref{Eq:avER}) is 4\% to 25\%  in the energy interval $(0.1,10)$ EeV; for Sibyll and EPOS models the divergence is not greater than 2\%.

Although the superposition approximation is not necessarily needed for our method to be applicable, it is convenient to accept it for the simplicity of conclusions. So, we can use the results of Sibyll2.1 or EPOSv1.99 models, together with equations (\ref{Eq:avER}) and (\ref{Eq:Dxm}) within the intervals $E\in(0.1,10)$ EeV, $A\in(1,56)$, where the discrepancy is less or equal to 2\%.

A straightforward approach to estimation of a difference in average mass composition is to compare  $\overline{x_m}$ of two samples of showers in the fixed energy interval. In the case of normal distributions with equal dispersions this is $t$-test mentioned above. If there is a statistically significant difference $\Delta x_m$ then one can certainly conclude that it is owing to the difference in average masses of samples, $\Delta\ln A$.

Actually, this approach is inefficient because of $\overline{x_m}$ rising with energy, and mass-dependent RMS deviation $\sigma(x_m)$. The most stringent restriction is caused by energy dependence: the two samples should be within a narrow energy interval, where few showers are detected with present-day arrays. For instance, the maximum number of EAS events detected in the $\lg E$ interval of width 0.2, where $x_m$ observations are available, is 1287 for the PAO data (\cite{PAOxm}, $\lg E\in(18.0,18.2)$), 171 for the HiRes data (\cite{HiResXm}, $\lg E\in(18.0,18.2)$), and 68 for the TA data (\cite{TAxm}, $\lg E\in(18.6,18.8)$).

\begin{table}[t]
\caption{Maximal resolution of the differences in average $x_m$ and UHECR mass, $\Delta\ln A$, with Student's $t$-test, using available data from EAS arrays.
}
\center\begin{tabular}{rcccc}&&&&\\
\hline
Detector & $\sigma(x_m)$ & $\Delta x_m$ & $\Delta\ln A$ &\\\cline{4-5}
         &      g/cm$^2$ &  g/cm$^2$    & Sibyll2.1 & EPOSv1.99\\
\hline
     PAO &            56 &          7.2 &      0.29 & 0.27\\
   HiRes &            52 &         20.0 &      0.81 & 0.76\\
      TA &            52 &         32.1 &      1.30 & 1.21\\
\hline
\end{tabular}\end{table}

In Table 1 the estimations of the majorant resolution are given for models and datasets. Here, the $t$-test is assumed applicable and the mass dependence of $x_m$ dispersion is neglected, so the real resolution should be worse. The maximal number of events detected in $\Delta\lg E=0.2$ intervals are divided equally between samples; $\Delta x_m$ resolvable by the $t$-test is calculated as in Figure \ref{fig:Comparison}. In order to estimate $\Delta\ln A$, equation \ref{Eq:Dxm} is used.

Wilcoxon test results are comparable or slightly weak. The experimental values of $\partial x_m/\partial\ln E$ are influenced by the possible changes of mass composition with energy; so one has to use the model simulation of $D_{ER}$ with fixed mass, in order to estimate the resolution.

Assessment of the resolution is ambiguous. In this paper, we assume the resolution sufficient if 10\% flux of $Fe$ nuclei is resolved in the background consisting of protons. In this context, the maximal resolution in $x_m$ is comparable with experimental uncertainties, the resolution of the average mass differences is insufficient, or hardly sufficient in the PAO case. Only future arrays with considerably larger aperture may improve the efficiency of this method.

Another approach should be used, that is applicable to all sorts of $x_m$ distributions with energy- and mass-dependent parameters. WSRT is a promising method which can be adapted to the case. In order to use WSRT for the analysis of EASs arriving from different celestial regions, we have to compare the rank sum of $x_m$ in the series of events. Due to the limited number of events available at the highest energies of interest, we should extend the boundaries of energy interval from which to select showers.

To do so, the paired samples can be used, in which every EAS from one sample has its counterpart in another sample with the same or closest energy. In this case, the two samples have the same $\overline{x_m}$ if the original mass compositions are identical, in spite of energy-dependent $\overline{x_m}$ and $\overline{A}$. On the other hand, if there is a difference in the average mass of original populations, then $\Delta x_m$ should exist and is approximated by (\ref{Eq:avER}).

With known efficiency of WSRT in resolving $\Delta x_m$ (Figure \ref{fig:Comparison}) and model simulations of $D_{ER}$, we can estimate $\Delta\overline{\ln A}$ for the observational number of showers. Namely, simulation results with the CORSIKA code of $\partial x_m/\partial\ln A$ for Sibyll (25, 24 and 23 g/cm$^2$), EPOS (27, 26 and 26 g/cm$^2$) models at energies (1, 10 and 55 EeV) \citep{Kampert} are used to relate $\Delta x_m$ and $\Delta\ln A$ in equation \ref{Eq:Dxm}.

In Table 2, the results for the PAO FD real data \citep{PAOxm}, and the estimated number of UHECRs ($E>55$ EeV) to be detected using JEM-EUSO during 5 years of orbiting the earth onboard the ISS \citep{EUSO} are shown. Duty cycle 0.19 and cloud impact 0.7 factors are accepted for JEM-EUSO exposure. The resolution in $x_m$ is assumed to be 120 g/cm$^2$.

The number of events is only sufficient in the PAO FD case for the threshold energy 1 EeV. The data of JEM-EUSO will not be convenient for this kind of analysis, owing mainly to poor resolution in the depth of shower maximum.

To increase the number of events under analysis at the highest energies, the data of the surface detectors (SD) can be used. This possibility will be discussed in the next section.

\begin{table}[t]
\caption{Minimal difference in average mass of UHECRs, $\Delta\overline{\ln A}$, resolvable by WSRT based on the data observed by PAO fluorescence detectors and JEM-EUSO planned statistics.}
\center\begin{tabular}{rrccc}&&&&\\
\hline
                       &           & $\Delta\overline{\ln A}$&\\\cline{3-4}
            Experiment & $N_{obs}$ & Sibyll2.1 & EPOSv1.99 \\
\hline
       PAO ($E>1$ EeV) &      6744 &      0.13 & 0.12\\
      PAO ($E>10$ EeV) &       339 &      0.61 & 0.56\\
 JEM-EUSO ($E>55$ EeV) &           &       & \\
            nadir mode &       550 &      1.07 & 0.98\\
           tilted mode &      1800 &      0.59 & 0.54\\
\hline
\end{tabular}\end{table}


\subsection{A method based on the age variance of air showers initiated by different primary particles}
There are some shower parameters measured by the surface detectors of the EAS arrays, sensitive to the nuclear composition of the primary particle: muon content, shower front curvature, etc. It seems that one of the most appropriate parameters is the shower age.
It is related explicitly to $x_m$\footnote{for the fixed zenith angle}, hence, to the primary mass, and can be estimated in each shower using the universal relation with lateral distribution (LD) parameters, measured by the SDs, in particular, the slope, $\eta$, of the charged particles LD \citep{IJMP}. Only electromagnetic component detectors are needed in the surface stations in this case, so the method can be applied to arrays with no muonic and other component detectors.

As in the previous section, having two paired samples of EAS ages one can apply WSRT in order to decide whether there is an appreciable difference in the mean age of the showers in samples $S$ and $T$, or whether these samples are drawn from the same distribution (the null hypothesis). To do so, one has to compare the ranks in samples of ages. Consulting with statistical tables about the deviation of statistic from the expected value, one can accept or reject the null hypothesis at the significance level specified.

Using the age variance instead of $\Delta x_m$ in EASs initiated by different primary particles, one has an additional variable to fix: the zenith angle. So, the paired samples should be selected with close or equal energies and zenith angles.
This can be done as follows: initially, one has two subsets of EASs arrived from different sources, and presumably, with different masses. For each shower from the minor subset, a counterpart shower should be found in another subset with equal or closest $\theta$ and $E$, then ages of these EASs should be collected in corresponding samples $S$ and $T$. Selected showers should be removed from subsets. Repeating operations till the end of the minor subset, one assembles a pair of samples of size $N$ congruous, on average, at $N\gg1$, to the circumstances of zenith angle and energy.

A relation between shower age and LD slope can be measured experimentally by PAO and TA fluorescence and surface detectors working together in the same showers. Meanwhile, an estimation can be used, derived by \cite{IJMP} using the CORSIKA code with implemented SIBYLL2.1/UrQMD models.

A minimal mass difference, resolvable by WSRT applied to age samples, is limited by uncertainty in age estimation and zenith angle. Neglecting the uncertainty of $\theta$ measurement in PAO data ($d\cos\theta/\cos\theta<0.01$), and using the number of EAS events detected during the period between 1.01.2004 and 31.12.2010 \citep{PAO}, we have the estimation of the minimal difference in average mass, resolvable by WSRT applied to PAO SD data (Table 3). Here, the minimal difference in shower age resolvable by the method is used, which in turn, is related to the difference in $x_m$: $|\Delta\tau/\tau|=|\Delta x_m/x_m|$ for a fixed zenith angle. $\Delta x_m$ is calculated applying the test for a given number of EAS events in samples $\sim N_{obs}/2$; average values of $x_m$ are given by the approximation of the PAO data \citep{PAOxm}.

\begin{table}[t]
\caption{Estimation of the minimal difference in the shower age, $\Delta\tau$, and average mass, $\Delta\overline{\ln A}$, of UHECRs observed by PAO surface detectors, resolvable by WSRT.}
\center\begin{tabular}{rrcccc}&&&&&\\
\hline
               &       &              & $\Delta\overline{\ln A}$&\\\cline{4-5}
$E_{thr}$, EeV &   $N_{obs}$ & $\Delta\tau$ & Sibyll2.1 & EPOSv1.99 \\
\hline
             3 & 63376 &       0.0015 &      0.04 & 0.03\\
            10 &  4790 &       0.0061 &      0.16 & 0.15\\
            25 &   608 &       0.0102 &      0.28 & 0.26\\
\hline
\end{tabular}\end{table}

Comparison of Tables 2 and 3 shows that, in general, the number of events detected with PAO SDs is sufficient (contrary to FDs with reduced duty cycle) to apply WSRT to the shower age and mass variance of EAS primaries, in the energy range above 10 EeV, and may be applicable at energies above 25 EeV. Concerning future arrays, large apertures, sufficient resolution in $x_m$ and/or $\tau$, and additionally, the presence of the surface detectors would be essential conditions of the applicability of the method.


\section{Application of the method to analysis of the Yakutsk array data}
In this section, the trial run of the method is realized with the Yakutsk array data.


\subsection{The Yakutsk array}
The Yakutsk array detects EAS of cosmic rays in the energy interval from 1 PeV to 100 EeV. The array is located at $61.7^0$N,$129.4^0$E, 105 m above sea level ($1020$ g cm$^{-2}$). It consists of 71 surface and 6 underground scintillation detectors of charged particles (electrons and muons), and 49 detectors of the air Cherenkov light. The total area covered by detectors with 500 m separation is $\sim10$ km$^2$. The array has been operating since 1970, and approximately $10^6$ showers of the primary energy above about $10^{15}$ eV have been detected. The highest energy event ($E\sim100$ EeV) detected was 7.05.1989 with an axis within the array area, at zenith angle $\theta=59^0$. More extended description of the array and results obtained can be found in~\cite{CRIS,NJP,EFe}.

\begin{figure}[t]
\includegraphics[width=0.85\columnwidth]{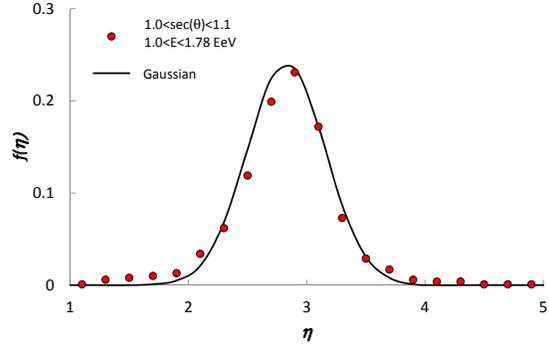}
\caption{A distribution of the LD slope, $\eta$. 3058 EAS events detected with the Yakutsk array (points) are selected in the energy and zenith angle intervals, defined in the upper left corner. Normal approximation with the same mean value and dispersion is plotted.}
\label{fig:SlopeSpread}
\end{figure}


\subsection{Slope of the lateral distribution function of charged particles in EAS}
An air shower cascading higher in the atmosphere (`old' shower, $\tau>1$) has a broad and flat lateral distribution of secondary particles at the observational level, while a `young' one ($\tau<1$) has a steep LD. It was shown previously that the LD parameters (slope of the Cherenkov light and charged particles lateral distributions, etc.) can be used as
indicators of the shower age \citep{CherSlope,Schmidt,IJMP}.

In this work, the LD slope of charged particles detected with scintillators of the Yakutsk array is used to estimate the shower age in the given energy and zenith angle intervals. Cherenkov light data are not used because of the small sample size of showers having this kind of signal detected. The same reason concerns the muon detectors data of the array.

The dataset used to analyze the slope parameter consists of EAS events collected during the period 1974 -- 2004, with energies from $1$ to $100$ EeV, zenith angles $\theta<50^0$ and axes within the array area. Inclined events beyond $50^0$ are rejected because of substantial fraction of muons in the distribution of charged particles measured. In order to estimate LD slope of each shower in a set, additional selection criteria were applied: i) at least 4 stations in the core distance interval $r\in(200,1000)$ m should have particle density above a threshold; and ii) the slope calculated using the least square method should be in the interval $\eta\in(-8,0)$. A total number of events survived after rejections is 19600.

A distribution of slopes in the narrow interval of energy and zenith angle is illustrated in Figure \ref{fig:SlopeSpread}. Normal approximation is rejected  by the Pearson's $\chi^2$ test because of the test-statistic equal to $\chi^2=1697.8$, while 23.2 is expected at the significance level $P_{crit}$ with 10 degrees of freedom.

\begin{figure}[t]
\includegraphics[width=\columnwidth]{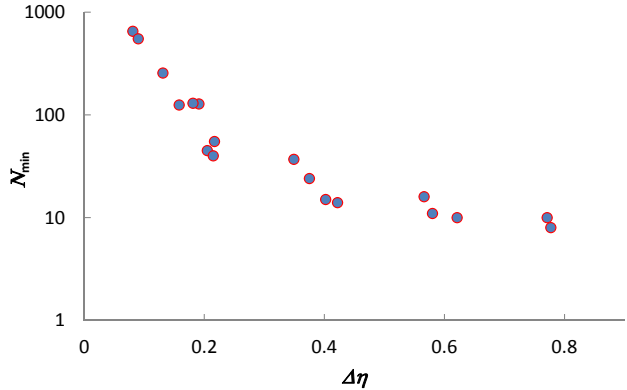}
\caption{Efficiency of WSRT applied to paired LD slope samples. The minimal sample size, $N_{min}$, sufficient to reject the null hypothesis at $P_{crit}$ when a difference in the mean slope of samples, $\Delta\eta$, is given.}
\label{fig:Power}
\end{figure}

The same procedure as in Section 2.4 is used to apply the Wilcoxon test to paired samples, except for the shower age replaced by the LD slope here. Sample $S$ consists of showers with arrival directions in supergalactic (SG) `pancake' or `dumbbell' structure around the SG plane \citep{Lahav}. Sample $T$ consists of all other showers. A hypothesis tested is that UHECR sources in SG pancake emit nuclei, whereas from other (distant) sources protons arrive. Our task is to ascertain if there is an appreciable difference in average mass of UHECRs in two samples.

In order to reveal the reliability bounds of WSRT in our particular case, we have applied a procedure to pair of EAS event samples selected in the narrow primary energy bin $E\sim1$ EeV\footnote{where CRs are presumably of Galactic origin and homogeneous in composition as a result of confinement in magnetic fields}, arrival directions within the whole sky, but in adjacent zenith angle intervals, so that paired showers are of the same energy, and with given difference in $\sec\theta$. This results in the same $x_m$ but different ages of paired EASs. Due to the universality of EAS, the average shower age is connected with the LD slope \citep{IJMP}.

Shifting adjacent zenith angle intervals, we can assign the difference in average LD slopes of samples. Then, we have to determine the minimal subsample size, $N_{min}$, sufficient to reject the null hypothesis that there is no difference in our source samples of slopes, by applying Wilcoxon test to paired subsamples. It is the efficiency of WSRT in the case of paired LD samples. In Figure \ref{fig:Power} this limit is shown as a function of the given difference in average slopes. Experimental errors in charged particle densities measured by scintillators result in uncertainties of differences, $\Delta\eta$, and in dispersion of points in the plot. The distribution of signed rank sum in the case of null hypothesis is calculated extracting two paired subsamples of size $N_{min}$ from a single sample of LD slopes.

Although in Optical Redshift Survey and IRAS 1.2-Jy redshift survey data the density contrast in the local galaxy density field is aligned along SG axes with radius 40 $h^{-1}$ Mpc and thickness of 20 $h^{-1}$ Mpc \citep{Lahav}, we have used three variants of the pancake angular boundaries in supergalactic latitudes: $b_{SG}<15^0/30^0/45^0$ in order not to miss the possible correlation of UHECRs with the SG plane proposed by \cite{Stanev}.

\begin{figure}[t]
\includegraphics[width=0.8\columnwidth]{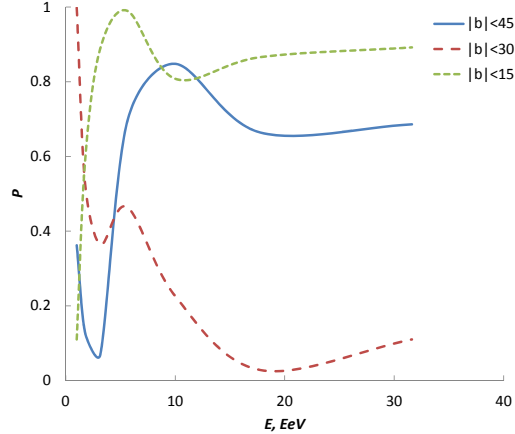}
\caption{The probability of two LD slope samples to be drawn from the same distribution.}
\label{fig:Pancake}
\end{figure}

The results are shown in Figure \ref{fig:Pancake}. In all cases we cannot reject the null hypothesis: there is no appreciable difference in LD slopes of EAS samples. Only at $E\sim 20$ EeV, $b_{SG}<30^0$ is there the minimum of a probability, but it is greater than $P_{crit}$.

In Table 4 the minimal differences in EAS age-related parameters are given, resolvable by WSRT, with the sample sizes determined by the threshold energy. Simulation results concerning relations among the LD slope and the shower age and $x_m$, derived with the Sibyll model, are used in this case \citep{IJMP}. A conclusion to be drawn is that the Yakutsk array data are too scanty to distinguish possible difference in mass composition of CRs, either at the threshold energy 10 EeV or 26 EeV, corresponding to C and Fe, respectively\footnote{Here, the rigidity 1 EeV is assumed as a very minimal limit to charged particles not confined in the Galaxy\citep{Horandel}.}. Instead, the data are sufficient, and can be used to search for variations in the composition of galactic CR at energies below 1 EeV.

\section{Conclusions}

A new method is developed, combining the analysis of UHECR arrival directions with the statistical test of paired EAS samples. It is applicable in cases when sky regions can be separated, where CR mass compositions are presumably different. The most straightforward application of the method lies at high energies ($E>10$ EeV)  where magnetic deflections are smallest and are not expected to completely isotorpize the CR sky distribution.

\begin{table}[b]
\caption{Estimation of the minimal difference in EAS parameters detected with the Yakutsk array, resolvable by WSRT. $N$ is the number of CRs arrived at $|b_{SG}|<30^0$; differences: in LD slope, $\Delta\eta$, in the shower age, $\Delta\tau$, in average CR mass, $\Delta\ln A$.}
\center\begin{tabular}{rrccc}&&&&\\
\hline
$E_{thr}$, EeV &  $N$ & $\Delta\eta$ & $\Delta\tau$ & $\Delta\ln A$\\
\hline
           1.0 & 9180 &        0.014 &        0.016 &          0.33\\
          10.0 &  114 &        0.120 &        0.133 &          3.10\\
          26.0 &   17 &        0.340 &        0.378 &          9.16\\
\hline
\end{tabular}\end{table}

The method is based on a non-parametric statistical test, WSRT, which does not depend on the populations fitting any parameterized distributions. Two algorithms are proposed: first, using measurements of the depth of EAS maximum position in the atmosphere; and second, relying on the age variance of air showers initiated by different primary particles.

The efficiency of WSRT is estimated with the distributions of shower maximum and the slope of lateral distribution in EAS. It is shown that the efficiency of Wilcoxon and Student tests are approximately equal in the case of normal distribution of the variable.

It is also shown that the data amount, concerning $x_m$ measurements with present-day EAS arrays, and even planned for JEM-EUSO telescope, is not sufficient to distinguish a 10\% flux of iron nuclei from the proton background at the significance level 0.01. However, measurements of EAS parameters related to the shower age with the surface detectors, specifically in the PAO experiment,  provide the bulk of the data able to reveal such a flux.

A trial run of the test is performed with the Yakutsk array data in order to search for a possible difference in the mass of CRs arriving from the supergalactic plane, and a complementary region. No significant difference is found in the LD slope parameter of the two subsets of EAS events detected with different threshold energies.

Using the correlation between the LD slope and the shower age, and maximum position provided by the Sibyll model simulations of EAS initiated by different primary particles, the estimations are given of minimal differences in these parameters, as well as the average mass of CRs, resolvable by the WSRT. The number of EAS events detected with the Yakutsk array is found to be insufficient to indicate any difference in masses of CRs, in the energy range above 1 EeV. Instead, it can be used to search for variations of galactic CR composition below this energy.

The method developed is quite general and can be applied at other energies and to the various EAS parameters measured at existing and future observatories.

\acknowledgments

The author is grateful to the Yakutsk array staff for data acquisition and valuable discussions. The Yakutsk array experiment is funded by the Russian Ministry of Education and Science and by RAS; this work is partially supported by RFBR grants \#11-02-00158 and \#11-02-12193.

\end{document}